\documentclass{aip-cp}

\usepackage[numbers]{natbib}
\usepackage{rotating}
\usepackage{graphicx}

\begin{document}

% Title portion
\title{New Mathematical Models of GPS Intersatellite Communications in the Gravitational Field of the Near-Earth Space}

\author[aff1]{Bogdan G. Dimitrov\corref{cor1}}

\affil[aff1]{ Institute of Nuclear Research and Nuclear Energetics, Bulgarian Academy of Sciences,\\ 72 Tzarigradsko~shaussee~Blvd., BG-1784 Sofia, Bulgaria}
\affil[aff2]{ Institute for Advanced Physical Studies, New Bulgarian University, 21 Montevideo Blvd., BG-1618 Sofia, Bulgaria}
\corresp[cor1]{Corresponding author: dimitrov.bogdan.bogdan@gmail.com}

\maketitle

\begin{abstract}
 {\itshape} Several space missions such as GRACE, GRAIL, ACES and others rely on intersatellite communications (ISC) between two satellites at a large distance one from another. The main goal of the theory is to formulate all the navigation observables within the General Relativity Theory (GRT). The same approach should be applied also to the intersatellite GPS-communications (in perspective also between the GPS, GLONASS and Galileo satellite constellations). In this paper a theoretical approach has been developed  for ISC between two satellites moving on (one-plane)  elliptical orbits based on the introduction of two gravity null cones with origins at the emitting-signal and receiving-signal satellites. The two null cones  account for the variable distance between the satellites during their uncorrelated motion.  This intersection of the two null cones gives  the space-time interval in GRT. Applying some theorems from higher algebra, it was proved that this space-time distance can become zero, consequently it can be also negative and positive.  But in order to represent the geodesic distance travelled by the signal, the space-time interval has to be "compatible" with the Euclidean distance. So this "compatibility condition", conditionally called "condition for ISC", is the most important consequence of the theory. The other important consequence is that the geodesic distance turns out to be the space-time interval, but with account also of the "condition for ISC". The geodesic distance is proved to be  greater than the Euclidean distance - a result, entirely based on the "two null cones approach" and moreover, without any use of the Shapiro delay formulae. Application of the same higher algebra theorems shows that the geodesic distance cannot have any zeroes, in accord with being greater than the Euclidean distance. The theory also puts a restriction on the eccentric anomaly angle E=45.00251 [deg], which is surprisingly close to the  angle of disposition of the satellites in the GLONASS satellite constellation (the Russian analogue of the American GPS) - 8 satellites within one and the same plane equally spaced at 45 deg. The approach is the first step towards constructing a new, consistent relativistic  theory of ISC between moving  satellites on different space-distributed Kepler orbits, which is a much more complicated problem not being solved for the moment.
\end{abstract}

% Head 1
\section{INTRODUCTION}
Currently GPS (Global Positioning System) technologies have developed rapidly due to the wide
implementation of atomic clocks \cite{Ludlow:AAB1}.

In the past $10$ years the problem about GPS satellite-ground station
communications has been replaced by the problem about autonomous navigation
and intersatellite communications (ISC) (links), which has been mentioned yet
in 2005 in the monograph \cite{Moritz:C27}. Autonomous navigation means that
generations of satellite Block II F (replenishment) and Block III
satellites have the capability to transmit data between them via
intersatellite cross-link ranging \cite{Xie:G1} and thus, to ensure navigation control and data
processing without commands from Earth stations in the course of six months.
 It is important that next generation space missions will attain submillimeter
precision of measuring distances beyond $10^{6}$ meters by means of ultrashort
femtosecond pulse lasers. The theoretical description of such
measurements is inevitably related with General Relativity Theory (GRT).

The theory of intersatellite communications (ISC) is developed in the series
of papers by S. Turyshev, V. Toth, M. Sazhin \cite{Turysh1:AAB62}, \cite{Turysh2:AAB63}
and S. Turyshev, N. Yu, V. Toth \cite{Turysh3:AAB64}. This theory  concerns the space missions
GRAIL (Gravity Recovery and Interior Laboratory), GRACE- FOLLOW-ON
(GRACE-FO - Gravity Recovery and Climate Experiment - Follow On) mission and
\ the Atomic Clock Ensemble in Space (ACES) \ experiment \cite{ACES:ACES1},
\cite{Salom:AAB51}  on the International Space Station (ISS). However, it should be stressed that
all these missions are realized by means of low-orbit satellites (about $450$ $km$ above the Earth) and
the distance between the satellites is about $220$ $km$, while the satellites from the GPS, GLONASS and
the Galileo constellations are on a much higher orbit (from $19140$ $km$ for GLONASS and ranging to $26560$ $km$
for GPS, even higher). Consequently, the distance between the satellites from one or from different constellations cannot be
determined with the accuracy ($1$ micrometer), typical for the GRAIL and GRACE missions, accounting also of the relative motion between the satellites.

In all these papers, the basic theoretical instrument for calculating the propagation time
for the signal is the Shapiro delay formulae. If the coordinates of the emitting and of the receiving
 satellite are correspondingly $\mid x_{A}(t_{A}%
)\mid=r_{A}$ and $\mid x_{B}(t_{B})\mid=r_{B}$, and $R_{AB}=$ $\mid
x_{A}(t_{A})-x_{B}(t_{B})\mid$ is the Euclidean distance between the signal - emitting satellite
and the signal - receiving satellite, then from the null cone equation, the signal
propagation time $T_{AB}=T_{B}-T_{A}$ between two space points can be expressed by the known formulae
\cite{Petit:AAB21} (see  also the review article
\cite{Interf:CA7A2B1} by Sovers, Fanselow and Jacobs on VLBI radio
interferometry)
\begin{equation}
T_{AB}=\frac{R_{AB}}{c}+\frac{2GM_{E}}{c^{3}}\ln\left(  \frac{r_{A} \nonumber  \\
+r_{B}+R_{AB}}{r_{A}+r_{B}-R_{AB}}\right) \ \ ,
\label{AA19}%
\end{equation}
where $GM_{E}$ is the geocentric gravitational constant and $M_{E}$ is the
Earth mass. We shall denote also by $t$=TCG the Geocentric Coordinate Time (TCG). The second term in
formulae Eq. (\ref{AA19}) is the Shapiro time delay, \ accounting for the signal
delay due to the curved space-time. In other words, due to the time delay, a propagating signal in
the gravitational field will travel a greater distance - this important fact from GRT will be
confirmed also by the applied new theoretical approach.

The key objective of the paper is to construct a mathematical formalism for the exchange of signals in the gravitational field of the Earth between GPS-satellites, which are not stationary with time, but are moving on one-plane elliptic orbits. A new theoretical approach in this paper will be the introduction of two gravitational null cones with origins at the signal-emitting and signal receiving satellites. The two null cones signify a transition to the moving reference systems of the two satellites, parametrized by the Kepler parameters (the eccentricities, the semi-major axis and the eccentric anomaly angles) for the case of plane elliptic motions. The changing Euclidean distance $R_{AB}$ between the satellites in fact means that the known formulae for the Shapiro time delay cannot be applied because it presumes non-moving emitters and receivers in the gravitational field around a massive body.

At the same time, there is also a concrete experimental situation, related to the RadioAstron
interferometric project, where the baseline distance (which is in fact $R_{AB}%
$) is changing. RadioAstron is a ground-space interferometer,
consisting of a space radio telescope (SRT) with a diameter $10$ meters,
launched into a highly elongated and perturbed orbit \cite{Sazhin:AAB42}, and a
ground radio telescope (GRT) with a diameter larger than $60$ $m$. The
baseline between the SRT and the GRT is changing its length due to the
variable parameters of the orbit. In this conjunction of SRT and a GRT, commonly called VLBRI
(Very Large Baseline Radio Interferometry) \cite{Kardash:AAB43}, the SRT time
turns out to be undefined due to the changing delay time, which is a
difference between the time of the SRT and the time of the Terrestrial
Station (TS). This might mean that the commonly accepted formulae for the
Shapiro time delay might not account for the relative motion between the SRT
and the GRT.

This report has the purpose just to outline the basic topics
and main conclusions in a more extensive and detailed theoretical research,
performed in the paper \cite{BOGDAN:PRD}. In this paper, the investigation is
restricted to intersatellite communications between satellites on one orbital plane.
This represents also a real situation from an experimental point of view. For example,
the satellites of the Galileo constellation (the European satellite system) are situated
on three orbital planes with nine-equally spaced operational satellites in each plane, which are in nearly
circular (i.e. slightly elliptical) orbits  with semi-major axis of $29600$\ km and a period of about $%
14 $\ hours \cite{Xu:C26}. The Russian Global Navigation Satellite System $%
GLONASS$, managed by the Russian Space Forces,
consists of $21$ satellites in three orbital planes (with three non-orbit
spares). Each satellite operates in nearly circular (again, slightly elliptical) orbits
with semi-major axis of $25510$ km, and the satellites within the same orbital plane are
equally spaced by $45$ degrees. The (USA) $%
GPS$ satellite constellation consists of $24$ operational satellites,
deployed in six evenly spaced planes ($A$ to $F$) with $4$ satellites per
plane and an inclination of the orbit $55$ degrees \cite{Moritz:C27}.

The theory in the paper \cite{BOGDAN:PRD} is just the first step
towards  constructing a theory for propagation of signals between moving satellites
on different, space-oriented orbits. The main motivation comes from the
requirement that the \textit{Global Navigation Satellite System} ($GNSS$),
consisting of $30$ satellites and orbiting the Earth at a height of $23616$
km, should be interoperable with the other two navigational systems $GPS$
and $GLONASS$ \cite{Soffel:C25}. This means that satellites on different orbital
planes should be able to exchange signals between each other, accounting for
the action of the gravitational field on the propagation of the signal. One more
motivation for developing such a theory is that a combined GNSS  of $75$ satellites from
the GPS, GLONASS and the Galileo constellations may increase greatly the visibility of the
satellites, especially in critical areas such as urban canyons \cite{Xu:C26}.

\section{TWO GRAVITY NULL CONES ACCOUNTING FOR THE VARIABLE BASELINE DISTANCE $R_{AB}$}
Let the emission time $T$ is the time coordinate in the metric
\begin{equation}
ds^{2}=0=g_{00}c^{2}dT^{2}+2g_{oj}cdTdx^{j}+g_{ij}dx^{i}dx^{j} \ \ .
\label{AAF3}%
\end{equation}
and the space coordinates are in fact the parametrization equations Eq. (\ref{C4AA1}),
where the space coordinates are parametrized in terms of the Keplerian (plane) elliptic orbital
parameters (semi-major axis $a$, eccentricity $e$ and eccentric anomaly
angle $E$)
\begin{equation}
x=a(\cos E-e)  , \  \   y=a\sqrt[.]{1-e^{2}}\sin E
\label{C4AA1}%
\end{equation}
for the (first) elliptic orbit. Further, the signal is intercepted by the receiver
of the second satellite and this signal is propagating on a (second) null cone
$ds_{(2)}^{2}=0$, where the metric tensor components $g_{00}$, $g_{oj}$ and
$g_{ij}$ are determined at a second space point $x_{2}$, $y_{2}$, parametrized
again by the equations Eq. (\ref{C4AA1}) in terms of new orbital parameters
$a_{2}$, $e_{2}$ and $E_{2}$. The peculiar and very important feature of the
newly proposed formalism in this paper is that we have two gravity null cones
$ds_{(1)}^{2}=0$ and $ds_{(2)}^{2}=0$ for the emitted and the received signal
(with time of emission $T_{1}$ and time of reception $T_{2}$) with cone
origins at the points $(x_{1},y_{1},0)$ and $(x_{2},y_{2},0)$. Also an
equation about the differential of the square of the Euclidean distance
$dR_{AB}^{2}$ is written, which now is a variable quantity. Thus, it can be
noted that the two propagation times $T_{1}$ and $T_{2}$ are no longer treated
in the framework of just one null cone equation (as is the case with the known
equation Eq. (\ref{AA19}) for the Shapiro time delay), but in the framework of two
gravitational null cone equations. These are the gravitational null cone metric
for the signal, emitted by the first satellite at the space point $(x_{1},y_{1},z_{1})$
\begin{equation}
ds_{1}^{2}=0=-(c^{2}+2V_{1})(dT_{1})^{2}+(1-\frac{2V_{1}}{c^{2}})\left((dx_{1})^{2}+(dy_{1})^{2}+(dz_{1})^{2}\right)
\label{ABC1}
\end{equation}
and the null cone metric for the second signal - receiving satellite
\begin{equation}
ds_{2}^{2}=0=-(c^{2}+2V_{2})(dT_{2})^{2}+(1-\frac{2V_{2}}{c^{2}})\left((dx_{2})^{2}+(dy_{2})^{2}+(dz_{2})^{2}\right) \ \ .
\label{ABC2}
\end{equation}
The changing positions (Euclidean
distance) between the satellites mean that the two four-dimensional null
cones have to be additionally intersected with the six-dimensional
hyperplane in terms of the variables $dx_{1}$, $%
dy_{1}$, $dz_{1}$, $dx_{2}$, $dy_{2}$, $dz_{2}$, which intersects the two
four-dimensional cones and depends also on the variable Euclidean distance
(meaning that $dR_{AB}^{2}\neq 0$)
\begin{equation}
dR_{AB}^{2}=2(x_{1}-x_{2})d(x_{1}-x_{2})+2(y_{1}-y_{2})d(y_{1}-y_{2})+2(z_{1}-z_{2})d(z_{1}-z_{2}) \ \ .
\label{ABC2A2}
\end{equation}%
This hyperplane equation is obtained after differentiation of the standard expression for the
Euclidean distance between the points $A$ and $B$
\begin{equation}
R_{AB}^{2}=(x_{1}-x_{2})^{2}+(y_{1}-y_{2})^{2}+(z_{1}-z_{2})^{2} \ \ .
\label{ABC2A1}
\end{equation}%
Consequently, the notions of Euclidean distance $%
R_{AB}$ and the propagation times $T_{1}$ and $T_{2}$ are closely related to
the intersection variety of the hyperplane with the two gravitational cones.
The derivation, the simultaneous solution of these three equations and some
physical consequences of the found solution
in terms of concrete numerical parameters for the GPS orbit are the main
objectives of this report.

\section{ANOTHER EXPRESSION FOR THE VARIABLE EUCLIDEAN DISTANCE FROM THE TWO INTERSECTING GRAVITATIONAL NULL CONES}
Our further aim will be to find a relation between the Euclidean distance
$R_{AB}$ (a notion from Newtonian mechanics) and the variables in the null
cone equations (notions from General Relativity Theory)). Making use of all the equations
in the preceding section, the following symmetrical relation is obtained
\[
dR_{AB}^{2}+S_{1}(E_{1},E_{2})dE_{1}+S_{2}(E_{1},E_{2})dE_{2}=
\]%
\begin{equation}
=P_{1}(E_{1})\frac{\left(  dT_{1}\right)  ^{2}}{dE_{1}}+P_{2}(E_{2}%
)\frac{\left(  dT_{2}\right)  ^{2}}{dE_{2}} \ \ .
\label{ABC17}%
\end{equation}
If the above equation is integrated as a differential equation in full derivatives (see the details in the paper
\cite{BOGDAN:PRD}), then the following symmetrical under interchange of the indices $1$ and $2$ expression can
be derived
\begin{eqnarray}
R_{AB}^{2}&=&\left(  -2e_{1}a_{1}^{2}\cos E_{1}-2e_{2}a_{2}^{2}\cos E_{2}\right) 
+\left(  2e_{2}a_{1}a_{2}\cos E_{1}+2e_{1}a_{1}a_{2}^{.} \cos E_{2}\right) \nonumber \\
&&+\frac{1}{2}\left(  e_{1}^{2}a_{1}^{2}\cos\left(  2E_{1}\right)  +e_{2}^{2}a_{2}^{2.}
\cos\left(  2E_{2}\right)  \right)  -2a_{1}a_{2}\cos E_{1}\cos E_{2}+2a_{1}a_{2}
\sqrt[.]{\left(  1-e_{1}^{2}\right)  \left(  1-e_{2}^{2}\right)}\sin E_{1}\sin E_{2} .
\label{ABC43}%
\end{eqnarray}
This is in fact the second 
representation for the square of the Euclidean distance. The first representation for
the function $R_{AB}^{2}$  is the
Euclidean distance $R_{AB}^{2}=(x_{1}-x_{2})^{2}+(y_{1}-y_{2})^{2}$,
  which can be presented as the following
symmetrical expression with respect to the two eccentric anomaly angles $E_{1}$ and $E_{2}$:
\begin{eqnarray}
R_{AB}^{2}&=&a_{1}^{2}+a_{2}^{2}+(a_{2}e_{1}
-a_{1}e_{2})^{2}-2a_{1}a_{2} \cos E_{1}\cos E_{2}
-2a_{1}a_{2}\sqrt[.]{\left(  1-e_{1}^{2}\right)  \left(  1-e_{2}^{2}\right)}
\sin E_{1}\sin E_{2} \nonumber \\
&&-a_{1}^{2}e_{1}^{2}\sin^{2}E_{1}-a_{2}^{2}e_{2}^{2}\sin^{2}E_{2}
+2a_{1}a_{2}(e_{2}\cos E_{1}+e_{1}\cos E_{2})
-2e_{1}a_{1}^{2}\cos E_{1}-2e_{2}a_{2}^{2}\cos E_{2}.
\label{ABC44}%
\end{eqnarray}
Relation Eq. (\ref{ABC43}) for the second representation has been found from the
intersection of the two null cones Eq. (\ref{ABC1}) and Eq. (\ref{ABC2}) and consequently,
 represents a General Relativity Theory (GRT) notion.
It is more correct to call it "a space-time interval", which according to
GRT can be either positive, negative or equal to zero. Although in Euclidean geometries
distances are strictly positive, "negative" distances are not prohibited and they are inherent to
the s.c. "Lobachevsky" geometries with negative scalar curvatures. In fact, an important clarification should be made:
GRT clearly defines what is a "gravitational null cone" and also a "space-time interval". GRT and also
Special Relativity Theory  do not give an answer to the problem: will the intersection
of two gravitational null cones again possess the property of the space-time interval, i.e. can it be
again positive, negative and zero? The investigation of the
"intersecting space-time interval" Eq. (\ref{ABC43}) in some partial simplified cases, but also in the general case will
give an affirmative answer to this problem. In other words, it will become evident that this "intersecting" interval
will again preserve the property of being positive, negative or null.

\section{THE COMPATIBILITY CONDITION FOR INTERSATELLITE COMMUNICATIONS}
There is nothing strange that the space-time interval can be different from the Euclidean distance and represents
a different expression from the Euclidean distance, since it is found also from another
equations (the gravitational null cone equations). But in any case, they denote one and the
same function denoted as $R_{AB}^{2}$. The only possibility for the compatibility of the two
representations is they to be equal both to zero or to be both positive.
However, as further it shall be explained, it is not obligatory to impose the requirement for the compatibility of the
two representations - the space-time interval Eq. (\ref{ABC43}) can be treated as an independent
notion from the Euclidean distance. But in the case of light or signal propagation, when the two
distances have to be compatible because the signal travels a macroscopic distance,
the two representations Eq. (\ref{ABC43}) and Eq. (\ref{ABC44}) have to be
set up equal. Then from the equality of the two representations Eq. (\ref{ABC43}) and Eq. (\ref{ABC44}) for
$R_{AB}^{2}$, one can obtain the following simple \ relation between the
eccentric anomalies, semi-major axis and the eccentricities of the two orbits
\begin{equation}
4a_{1}a_{2}\sqrt[.]{\left(  1-e_{1}^{2}\right)  \left(  1-e_{2}^{2}\right)
}\sin E_{1}\sin E_{2}
=a_{1}^{2}+a_{2}^{2}+(a_{2}e_{2}-a_{1}e_{1})^{2}-\frac{1}{2}\left(  e_{1}%
^{2}a_{1}^{2}+e_{2}^{2}a_{2}^{2}\right).
\label{ABC45}%
\end{equation}
This relation can be conditionally called "a condition for intersatellite
communications between satellites on (one - plane) elliptical orbit". It is
obtained as a compatibility condition between the large-scale, Euclidean distance
Eq. (\ref{ABC44}) and the space-time interval Eq. (\ref{ABC43}).

\section{POSITIVITY AND NEGATIVITY OF THE SPACE-TIME INTERVAL FOR THE CASE OF EQUAL ECCENTRIC ANOMALY ANGLES, ECCENTRICITIES AND SEMI-MAJOR AXIS}

The space-time interval for some
specific cases can be of any signs, while the situation will turn out to be
different for the "geodesic distance". It is remarkable that the
positivity of the geodesic distance will become evident when performing a
simple algebraic substitution of the condition Eq. (\ref{ABC45}) into formulae Eq. (%
\ref{ABC43}) and at the same time, this will be confirmed by the analysis of
a complicated algebraic equation of fourth degree.

The space-time interval Eq. (\ref{ABC43}) for the case
of equal eccentricities, semi-major axis and eccentric anomaly angles ($%
e_{1}=e_{2}=e$, $a_{1}=a_{2}=a$, $E_{1}=E_{2}=E$) can be represented as
\begin{equation}
R_{AB}^{2}=4a^{2}\sin ^{2}E.(1-e^{2})+a^{2}(e^{2}-2) \ \ .
\label{F1}
\end{equation}%
It can be noted that this space-time interval is positive for
\begin{equation}
\sin ^{2}E\geq \frac{2-e^{2}}{4(1-e^{2})} \ \ ,
\label{F2}
\end{equation}%
but it will be negative, when sign of the inequality will be the opposite one.

If the condition for intersatellite
communications Eq. (\ref{ABC45}) is taken into considerations, then for $e_{1}=e_{2}=e$, $%
a_{1}=a_{2}=a$, $E_{1}=E_{2}=E$ this condition gives the relation
\begin{equation}
4a^{2}(1-e^{2})\sin ^{2}E=2a^{2}-e^{2}a^{2}  \Longrightarrow \sin
E=\frac{1}{2}\sqrt[.]{\frac{\left( 2-e^{2}\right) }{\left( 1-e^{2}\right) }}% \ \ .
\label{ABC49A}
\end{equation}%
If substituted into the space-time interval Eq. (\ref{F1}), the above relation gives $%
R_{AB}^{2}=0$. This should be expected and in fact is a consistency check of
the calculations because for equal eccentricities, semi-major axis and
eccentric anomaly angles, the Euclidean distance Eq. (\ref{ABC44}) is equal to
zero. Then the compatibility condition Eq. (\ref{ABC49A}) should give also zero when
substituted in the formulae for the space-time interval.

\section{SPACE-TIME INTERVAL FOR THE CASE OF DIFFERENT ECCENTRIC ANOMALY ANGLES (NON-ZERO EUCLIDEAN DISTANCE) AND NEW PHYSICAL INTERPRETATION FOR THE EUCLIDEAN DISTANCE}

This is  the case of equal eccentricities and semi-major axis, but
different \ eccentric anomaly angles ($E_{1}\neq E_{2}$).
The space-time interval Eq. (\ref{ABC43}) can be written as
\begin{equation}
R_{AB}^{2}=e^{2}a^{2}-e^{2}a^{2}(\sin E_{1}+\sin E_{2})^{2}-2a^{2}\cos (E_{1}+E_{2}) \ \    .
\label{F6}
\end{equation}
After using some known formulaes for trigonometric functions, it can be proved that
\begin{equation}
\cos (E_{1}+E_{2})<\frac{e^{2}}{2}\Longrightarrow E_{1}+E_{2}>\arccos \left[
\frac{e^{2}}{2}\right] \ \  .  \label{F11}
\end{equation}
For the typical value $e=0.01323881349526$ of the eccentricity of the GPS orbit
(see the PhD thesis \cite{Gulklett:C4} of Gulklett), it can be obtained
\begin{equation}
E_{1}+E_{2}>89.994978993712 \ [\deg ] \ \ .
\label{F12}
\end{equation}%
This numerical result again confirms the consistency of the calculations, but
this can be proved if one returns again to the previous case of equal eccentricities,
semi-major axis and eccentric anomaly angles. Then the lower bound,  for which
$R_{AB}^{2}\geq 0$ for the case of the above typical GPS\ orbit eccentricity
is given by the limiting value $E_{\lim }$ for the eccentric anomaly angle
\begin{equation}
E_{\lim }=\arcsin \left[ \frac{1}{2}\sqrt{\frac{2-e^{2}}{1-e^{2}}}\right]=45.002510943228 \ [\deg ] \ \ .
\label{F4}
\end{equation}
However, formulae Eq. (\ref{F12}) should be valid also for the partial case of equal
eccentric anomaly angles. This means that twice the value of $45.002510943228$ should be
greater than the number $89.994978993712$ in the right-hand side of the same
formulae. This is indeed the case, consequently there is full consistency between the
two different cases. The number $E=45.002510943228  \ [\deg ]$ for the GPS orbit turns out to be important from the point of view of
the disposition of the satellites in the GLONASS satellite constellation, where eight satellites are situated
equally spaced at $45$ degrees within one and the same plane. For the GLONASS satellites with eccentricity of the
orbit $e=0.02$ (close to the eccentricity of the GPS orbit), the value for the eccentric
anomaly angle is obtained to be $E=45.00573$ $%
[\deg ]$, which is surprisingly close to the angle of equal spacing of the GPS satellites within one orbit.
For the Galileo constellation, the $9$ satellites per one plane are  equally spaced at $40$ degrees.
Consequently, the following question remains to be answered: what is the
role of the angle $45$ $[\deg ]$ in the \ GPS, GLONASS and (to a certain extent)-the Galileo intersatellite
communications?

Since the Euclidean distance is only positive, while the space-time interval for the intersection of
the two null cones can be either positive, negative or zero, the Euclidean distance can be affirmed
to represent a partial case of a more general case, related to the space-time distance. Thus, one can define the Euclidean
distance as a positive space-time distance, measured along the intersection
of two null four-dimensional gravitational null cones, attached to two moving
observers (on the emitting - signal satellite and on the receiving - signal
satellite).
\section{THE RESTRICTION ON THE ELLIPTICITY OF THE ORBIT, THE COMPATIBILITY CONDITION AND POSITIVE GEODESIC DISTANCE}
It was mentioned that the equality of the space-time and the Euclidean distance
gives the compatibility condition. But if the compatibility condition Eq. (\ref{ABC45})
is substituted into expression Eq. (\ref{ABC43}) for the space-time interval
, then Eq. (\ref{ABC43}) will give the "geodesic distance", denoted by $\widetilde{R}%
_{AB}^{2}$ in order to distinguish it from the space-time distance Eq. (\ref{ABC43}) and from the Euclidean
distance Eq. (\ref{ABC44}). Since the formulae for the geodesic distance is
\begin{eqnarray}
\widetilde{R}_{AB}^{2}&=&\frac{1}{2}(a_{1}^{2}+a_{2}^{2})+\frac{1}{2}\left(
a_{2}e_{2}-a_{1}e_{1}\right)  ^{2}+\frac{1}{4}\left(  a_{1}^{2}e_{1}^{2}%
+a_{2}^{2}e_{2}^{2}\right)
-\left(  2e_{1}a_{1}^{2}\cos E_{1}+2e_{2}a_{2}^{2}\cos E_{2}\right)  \nonumber \\
&&-\left(  e_{1}^{2}a_{1}^{2}\sin^{2}E_{1}+e_{2}^{2}a_{2}^{2}\sin^{2}%
E_{2}\right)  -2a_{1}a_{2}\cos E_{1}\cos E_{2}+
+2a_{1}a_{2}\left(  e_{2}\cos E_{1}+e_{1}\cos E_{2}\right),
\label{ABC46}%
\end{eqnarray}
it is easily verified that it is different from formulae Eq. (\ref{ABC43}) and formulae Eq. (\ref{ABC44}).
In the paper \cite{Klioner:AAB33} the geodesic distance  is defined as the distance, travelled by a light or radio
signal in the gravitational field. Since the formalism here is related to the method of "two null gravitational cones"",
by analogy with the Shapiro delay formulae Eq. (\ref{AA19}) , the following problem arizes: will the "geodesic" distance be greater than the
Euclidean one also for this special case ? In order to resolve this problem, let us write Eq. (\ref{ABC43})
not for the general case, but for the case of equal semi-major axis, eccentricities and eccentric anomaly angles.
Let us substract the square of the geodesic distance Eq. (\ref{ABC46}) from the square of the Euclidean
distance Eq. (\ref{ABC44})
\begin{equation}
\widetilde{R}_{AB}=\sqrt{R_{AB}^{2}+a^{2}\left[ 1-\frac{3}{2}e^{2}\right] }%
\ \ .  \label{ABC55A1}
\end{equation}
Note the interesting fact that this expression does not depend on the
eccentric anomaly angle $E$. In order to realize the importance of this result, one should take into account again
the compatibility condition Eq. (\ref{ABC45}) also for the case of equal semi-major axis, eccentricities and
eccentric anomaly angles. Taking into account the restriction for trigonometric .
functions  $\sin E\leq1$ , the following inequality can be obtained
\begin{equation}
\sin E=\frac{1}{2}\sqrt[.]{\frac{\left(  2-e^{2}\right)  }{\left(
1-e^{2}\right)  }}\leq1  \ \ ,
\label{ABC53A}%
\end{equation}
which is fulfilled for
\begin{equation}
e^{2}\leq\frac{2}{3}=0.816496580927726  \ \  .
\label{ABC54A}%
\end{equation}
Surprisingly, highly eccentric orbits (i.e. with the ratio $e=\frac
{\sqrt[.]{a^{2}-b^{2}}}{a}$ tending to one, where $a$ and $b$ are the great
and the small axis of the elliptical orbit) are not favourable for intersatellite
communications. For GPS\ satellites, which have very low eccentricity orbits
(of the order $e=0.01$) and for communication satellites on circular orbits
($e=0$), intersatellite communications between moving satellites can be
practically achieved.

    The more important consequence of the restriction Eq. (\ref{ABC54A}) $e^{2}\leq \frac{2}{3}$
on the value of the ellipticity of the orbit is that the second term under the
square root in Eq. (\ref{ABC55A1}) is positive. Due to this
\begin{equation}
\widetilde{R}_{AB}\geq R_{AB} \ \  ,
\label{ABC55A2}
\end{equation}%
which means that the geodesic distance, travelled by the signal is greater
than the Euclidean distance. This simple result, obtained by applying the
formalism of two intersecting null cones is a formal proof of the validity
of the Shapiro time delay formulae for the case of moving emitters and
receptors of the signals. Due to the larger geodesic distance, any signal in
the presence of a gravitational field will travel a greater distance and
thus will be additionally delayed. Consequently, this is also a confirmation of
the physical and mathematical consistency of the approach of two null gravitational
intersecting cones.

    It is also a remarkable feature that the geodesic distance can be proved to be greater than the
Euclidean one also in the general case of different eccentricities of the
two orbits, different semi-major axis and different eccentric anomaly angles. The details of the
not so complicated proof, based on algebraic manipulations,  can be seen in the paper \cite{BOGDAN:PRD}.

\section{THE SHUR THEOREM FROM HIGHER ALGEBRA APPLIED TO THE GENERAL CASE FOR THE SPACE-TIME INTERVAL AND THE GEODESIC DISTANCE}

The considerations in the previous sections for some partial cases clearly suggest that the space-time interval and the
geodesic distance have different properties - the space-time interval can be positive, negative or zero, while the geodesic
distance  can be only positive and non-zero, since it is greater than the Euclidean distance.

The basic and important problem is whether these properties of the space-time interval and the geodesic distance can be
extended to the general case. Previously, on the base of purely algebraic manipulations it was suggested that this generalization is
possible to be performed for the geodesic distance. However, it can be proved that the geodesic distance can be written in
the form of a fourth-order algebraic polynomial
\begin{equation}
g(y)=\overline{a}_{0}y^{4}+\overline{a}_{1}y^{3}+\overline{a}_{2}y^{2}+%
\overline{a}_{3}y+\overline{a}_{4}  \ \  ,
\label{E1}
\end{equation}%
where the notation $y=\sin^{2}E_{1}$ has been introduced and the coefficient functions are
complicated expressions, containing parameters and the other eccentric anomaly angle. Now it can be
noted that the variable $y$ should contain its values within the unit circle, in view of the properties of the
trigonometric functions.

Consequently, the problem about the geodesic distance being only positive is equivalent to the problem
of proving that the fourth-order algebraic equation $g(y)=0$ does not have any roots within the unit circle.

In a similar way, the space-time interval algebraic equation can be represented in the form
\begin{equation}
f(y)=a_{0}y^{4}+a_{1}y^{3}+a_{2}y^{2}+a_{3}y+a_{4}  \ \ .
\label{C25}
\end{equation}%
For this case, the physical property  of the space-time distance of being positive, negative and zero
results in a different algebraic formulation: the algebraic equation $f(y)=0$ should have roots and all
of them should remain within the unit circle $\mid y\mid <1$.

Both cases can be treated by means of the well-known Shur theorem from higher algebra, the formulation and the
proof of which are given in the monograph of the Bulgarian mathematician
Prof. Nikola Obreshkoff \cite{Obresh:AAB16AB}: \\Theorem:
The necessary and sufficient conditions for the polynomial of $n-$th degree
\begin{equation}
f(y)=a_{0}y^{n}+a_{1}y^{n-1}+....+a_{n-2}y^{2}+a_{n-1}y+a_{n}  \label{F27}
\end{equation}%
to have roots only in the circle $\mid y\mid <1$ are the following ones:

1. The fulfillment of the inequality
\begin{equation}
\mid a_{0}\mid >\mid a_{n}\mid  \ \ .  \label{F28}
\end{equation}

2. The roots of the polynomial of $(n-1)$ degree
\begin{equation}
f_{1}(y)=\frac{1}{y}\left[ a_{0}f(y)-a_{n}f^{\ast }(y)\right]
\label{F29}
\end{equation}%
should be contained in the circle $\mid y\mid <1$, where $f^{\ast }(y)$ is
the s.c. "inverse polynomial", defined as
\begin{equation}
f^{\ast }(y)=y^{n}f(\frac{1}{y}%
)=a_{n}y^{n}+a_{n-1}y^{n-1}+....+a_{2}y^{2}+a_{1}y+a_{0} \ \ .
\label{F30}
\end{equation}%
In case of fulfillment of the inverse inequality%
\begin{equation}
\mid a_{0}\mid <\mid a_{n}\mid  \label{F31}
\end{equation}%
the $(n-1)$ degree polynomial $f_{1}(y)$ (again with the requirement the
roots to remain within the circle $\mid y\mid <1$) is given by the
expression
\begin{equation}
f_{1}(y)=a_{n}f(y)-a_{0}f^{\ast }(y) \ \ .  \label{F32}
\end{equation}%
By means of this theorem, after some rather lengthy and cumbersome calculations, using
as well a modified version of this theorem, in the paper \cite{BOGDAN:PRD} it was proved that
the space-time fourth-order algebraic equation Eq. (\ref{C25}) indeed has all of its roots
contained in the unit circle $\mid y\mid <1$, while the geodesic fourth-order algebraic
equation Eq. (\ref{E1}) does not have any roots within this unit circle. Note also that the
treatment of both cases by means of this theorem is possible only because it possesses a
necessary and sufficient condition, which enables the application of the theorem with respect to a chain
of lower degree polynomials of third, second and finally - first-order. Thus, the physical
interpretation about the  physical notions of the "space-time" interval and the "geodesic distance" are
fully consistent with the mathematical results, which concern not only the partial cases, but also the general case
of non-equal eccentricities, semi-major axis and eccentric anomaly angles. The mathematical proof for the general case
by means of the Shur theorem was performed for the case of very small eccentricities of the orbit ($e\sim0.01$), which is consistent
with the real data for the GPS orbit.

% References

\nocite{*}
\bibliographystyle{aipnum-cp}

\end{document}